\documentclass{article}
\usepackage{graphicx}
\usepackage{amsmath}

\input{tcilatex}

\begin{document}

\title{\textbf{Extended Conformal Algebra and Non-commutative Geometry in Particle
Theory}}
\author{W. Chagas-Filho \\
Departamento de Fisica, Universidade Federal de Sergipe\\
SE, Brazil}
\maketitle

\begin{abstract}
We show how an off shell invariance of the massless particle action allows
the construction of an extension of the conformal space-time algebra and
induces a non-commutative space-time geometry in bosonic and supersymmetric
particle theories.
\end{abstract}

\section{Introduction}

\noindent Relativistic particle theory may be looked at as a prototype
theory for string theory and general relativity. Recently, there has been an
intense research activity on the connection between string theory and
non-commutative geometry. This is because it is believed that in quantum
theories containing gravity, space-time must change its nature at distances
comparable to the Planck scale. Quantum gravity has an uncertainty principle
which prevents one from measuring positions to better accuracies than the
Planck length [1]. These effects could then be modeled by a non-vanishing
commutation relation between the space-time coordinates. Many developments
in string and superstring theories were reached in this direction (see for
instance [2], [3], [4] and cited references), and it is now believed that
non-commutative geometry naturally arises from the string dynamics.

An approach to the study of the relations between string dynamics and
particle dynamics, using the concept of space-time symmetry as an
investigation tool, was presented in [5]. In particular, it was verified the
existence of a very special string motion in the high energy limit of the
theory. In this special motion, each point of the string moves as if it were
a massless particle. The existence of such a string motion agrees with what
is expected from gauge theory-string duality.

The above mentioned string motion opens the question of if we can find an
indication that non-commutative geometry should also naturally emerge from
the massless relativistic particle dynamics. In this work we consider this
question and show how we may use the special-relativistic ortogonality
condition between the velocity and acceleration to induce a new invariance
of the massless particle action. This induced symmetry may then be used to
construct an off shell extension of the conformal space-time algebra in four
dimensions and also to generate a transition to new space-time coordinates
which obey non-vanishing commutation relations. The extended conformal
algebra automatically reduces to the usual conformal algebra if the mass
shell condition is imposed. The new commutation relations obey all Jacobi
identities among the canonical variables, are preserved in the
supersymmetric theory, and reduce to the usual commutation relations if the
mass shell condition is imposed. The conclusion is that there exists an
inertial frame in which the uncertainty introduced by two simultaneous
position measurements is an off shell rotation in space-time. For clarity of
our exposition of the subject, next section briefly reviews the concept of
conformal vector fields.

\section{Conformal Vector Fields}

Consider the Euclidean flat space-time vector field 
\begin{equation}
\hat{R}(\epsilon )=\epsilon ^{\mu }\partial _{\mu }  \tag{2.1}
\end{equation}
such that 
\begin{equation}
\partial _{\mu }\epsilon _{\nu }+\partial _{\nu }\epsilon _{\mu }=\delta
_{\mu \nu }\partial .\epsilon  \tag{2.2}
\end{equation}
The vector field $\hat{R}(\epsilon )$ gives rise to the coordinate
transformation 
\begin{equation}
\delta x^{\mu }=\hat{R}(\epsilon )x^{\mu }=\epsilon ^{\mu }  \tag{2.3}
\end{equation}
The vector field (2.1) is known as the Killing vector field and $\epsilon
^{\mu }$ is known as the Killing vector. One can show that the most general
solution for equation (2.2) in a four-dimensional space-time is 
\begin{equation}
\epsilon ^{\mu }=\delta x^{\mu }=a^{\mu }+\omega ^{\mu \nu }x_{\nu }+\alpha
x^{\mu }+(2x^{\mu }x^{\nu }-\delta ^{\mu \nu }x^{2})b_{\nu }  \tag{2.4}
\end{equation}
The vector field $\hat{R}(\epsilon )$ for the solution (2.4) can then be
written as 
\begin{equation}
\hat{R}(\epsilon )=a^{\mu }P_{\mu }-\frac{1}{2}\omega ^{\mu \nu }M_{\mu \nu
}+\alpha D+b^{\mu }K_{\mu }  \tag{2.5}
\end{equation}
where 
\begin{equation}
P_{\mu }=\partial _{\mu }  \tag{2.6}
\end{equation}
\begin{equation}
M_{\mu \nu }=x_{\mu }\partial _{\nu }-x_{\nu }\partial _{\mu }  \tag{2.7}
\end{equation}
\begin{equation}
D=x^{\mu }\partial _{\mu }  \tag{2.8}
\end{equation}
\begin{equation}
K_{\mu }=(2x_{\mu }x^{\nu }-\delta _{\mu }^{\nu }x^{2})\partial _{\nu } 
\tag{2.9}
\end{equation}
$P_{\mu }$ generates translations, $M_{\mu \nu }$ generates rotations, $D$
generates dilatations and $K_{\mu }$ generates conformal transformations in
space-time. The generators of the vector field $\hat{R}(\epsilon )$ obey the
commutator algebra 
\begin{equation}
\lbrack P_{\mu },P_{\nu }]=0  \tag{2.10a}
\end{equation}
\begin{equation}
\lbrack P_{\mu },M_{\nu \lambda }]=\delta _{\mu \nu }P_{\lambda }-\delta
_{\mu \lambda }P_{\nu }  \tag{2.10b}
\end{equation}
\begin{equation}
\lbrack M_{\mu \nu },M_{\lambda \rho }]=\delta _{\nu \lambda }M_{\mu \rho
}+\delta _{\mu \rho }M_{\nu \lambda }-\delta _{\nu \rho }M_{\mu \lambda
}-\delta _{\mu \lambda }M_{\nu \rho }  \tag{2.10c}
\end{equation}
\begin{equation}
\lbrack D,D]=0  \tag{2.10d}
\end{equation}
\begin{equation}
\lbrack D,P_{\mu }]=-P_{\mu }  \tag{2.10e}
\end{equation}
\begin{equation}
\lbrack D,M_{\mu \nu }]=0  \tag{2.10f}
\end{equation}
\begin{equation}
\lbrack D,K_{\mu }]=K_{\mu }  \tag{2.10g}
\end{equation}
\begin{equation}
\lbrack P_{\mu },K_{\nu }]=2(\delta _{\mu \nu }D-M_{\mu \nu })  \tag{2.10h}
\end{equation}
\begin{equation}
\lbrack M_{\mu \nu },K_{\lambda }]=\delta _{\nu \lambda }K_{\mu }-\delta
_{\lambda \mu }K_{\nu }  \tag{2.10i}
\end{equation}
\begin{equation}
\lbrack K_{\mu },K_{\nu }]=0  \tag{2.10j}
\end{equation}
The commutator algebra (2.10) is the conformal space-time algebra in four
dimensions. Notice that the commutators (2.10a-2.10c) correspond to the
Poincar\'{e} algebra. The Poincar\'{e} algebra is a sub-algebra of the
conformal algebra. Let us now see how we can extend the conformal algebra,
and how this extended algebra is related to a non-commutative space-time
geometry.

\bigskip

\section{Relativistic Particles}

A relativistic particle describes is space-time a one-parameter trajectory $%
x^{\mu }(\tau )$. A possible form of the action is the one proportional to
the arc length traveled by the particle and given by 
\begin{equation}
S=-m\int ds=-m\int d\tau \sqrt{-\dot{x}^{2}}  \tag{3.1}
\end{equation}
In this work we choose $\tau $ to be the particle's proper time, $m$ is the
particle's mass and $ds^{2}=-\delta _{\mu \nu }dx^{\mu }dx^{\nu }$. A dot
denotes derivatives with respect to $\tau $ and we use units in which $\hbar
=c=1$.

Action (3.1) is obviously inadequate to study the massless limit of the
theory and so we must find an alternative action. Such an action can be
easily computed by treating the relativistic particle as a constrained
system. In the transition to the Hamiltonian formalism action (3.1) gives
the canonical momentum 
\begin{equation}
p_{\mu }=\frac{m}{\sqrt{-\dot{x}^{2}}}\dot{x}_{\mu }  \tag{3.2}
\end{equation}
and this momentum gives rise to the primary constraint 
\begin{equation}
\phi =\frac{1}{2}(p^{2}+m^{2})=0  \tag{3.3}
\end{equation}
We follow Dirac's [6] convention that a constraint is set equal to zero only
after all calculations have been performed. The canonical Hamiltonian
corresponding to action (3.1), $H=p.\dot{x}-L$, identically vanishes. This
is a characteristic feature of reparametrization-invariant systems. Dirac's
Hamiltonian for the relativistic particle is then 
\begin{equation}
H_{D}=H+\lambda \phi =\frac{1}{2}\lambda (p^{2}+m^{2})  \tag{3.4}
\end{equation}
where $\lambda (\tau )$ is a Lagrange multiplier. The Lagrangian that
corresponds to (3.4) is 
\begin{eqnarray}
L &=&p.\dot{x}-H_{D}  \notag \\
&=&p.\dot{x}-\frac{1}{2}\lambda (p^{2}+m^{2})  \TCItag{3.5}
\end{eqnarray}
Solving the equation of motion for $\ p_{\mu }$ that \ follows from (3.5)
and inserting the result back in it, we obtain the particle action 
\begin{equation}
S=\int d\tau (\frac{1}{2}\lambda ^{-1}\dot{x}^{2}-\frac{1}{2}\lambda m^{2}) 
\tag{3.6}
\end{equation}
The great advantage of action (3.6) is that it has a smooth transition to
the $m=0$ limit.

Action (3.6) is invariant under the Poincar\'{e} transformations 
\begin{equation}
\delta x^{\mu }=a^{\mu }+\omega _{\nu }^{\mu }x^{\nu }  \tag{3.7a}
\end{equation}
\begin{equation}
\delta \lambda =0  \tag{3.7b}
\end{equation}
Invariance of action (3.6) under transformation (3.7a) implies that we can
construct a space-time vector field corresponding to the first two
generators in the right of equation (2.5). These generators realize the
Poincar\'{e} algebra (2.10a-2.10c).

Now we make a transition to the massless limit. This limit is described by
the action 
\begin{equation}
S=\frac{1}{2}\int d\tau \lambda ^{-1}\dot{x}^{2}  \tag{3.8}
\end{equation}
The canonical momentum conjugate to $x^{\mu }$ is 
\begin{equation}
p_{\mu }=\frac{1}{\lambda }\dot{x}_{\mu }  \tag{3.9}
\end{equation}
The canonical momentum conjugate to $\lambda $ identically vanishes and this
is a primary constraint, $p_{\lambda }=0$. Constructing the canonical
Hamiltonian, and requiring the stability of this constraint, we are led to
the mass shell condition 
\begin{equation}
\phi =\frac{1}{2}p^{2}=0  \tag{3.10}
\end{equation}
Let us now study which space-time symmetries are present in this limit.
Being the $m=0$ limit of (3.6), action (3.8) is also invariant under
transformation (3.7). The massless action (3.8) however has a larger set of
space-time invariances. It is also invariant under the scale transformation 
\begin{equation}
\delta x^{\mu }=\alpha x^{\mu }  \tag{3.11a}
\end{equation}
\begin{equation}
\delta \lambda =2\alpha \lambda  \tag{3.11b}
\end{equation}
where $\alpha $ is a constant, and under the conformal transformation 
\begin{equation}
\delta x^{\mu }=(2x^{\mu }x^{\nu }-\delta ^{\mu \nu }x^{2})b_{\nu } 
\tag{3.12a}
\end{equation}
\begin{equation}
\delta \lambda =4\lambda x.b  \tag{3.12b}
\end{equation}
where $b_{\mu }$ is a constant vector. Invariance of action (3.8) under
transformations (3.7a), (3.11a) and (3.12a) then implies that the full
conformal field (2.5) can be defined in the massless sector of the theory.

It is convenient at this point to relax the mass shell condition (3.10).
This is because the massless particle will be off mass shell in the presence
of interactions [4]. We may now use the fact that we are dealing with a
special-relativistic system. Special relativity has the characteristic
kinematical feature that the relativistic velocity is always ortogonal to
the relativistic acceleration ( see, for instance, [7] ). Then, as a
consequence of the fact that action (3.8) is a special-relativistic action,
it is invariant under the transformation 
\begin{equation}
x^{\mu }\rightarrow \tilde{x}^{\mu }=\exp \{\beta (\dot{x}^{2})\}x^{\mu } 
\tag{3.13a}
\end{equation}
\begin{equation}
\lambda \rightarrow \exp \{2\beta (\dot{x}^{2})\}\lambda  \tag{3.13b}
\end{equation}
where $\beta $ is an arbitrary function of $\dot{x}^{2}$. We emphasize that
although the ortogonality condition must be used to get the invariance of
action (3.8) under transformation (3.13), this condition is not an external
ingredient in the theory. In fact, the ortogonality between the relativistic
velocity and acceleration is an unavoidable condition here, it is an
imposition of special relativity.

It may be pointed out that since the relativistic ortogonality condition is
used to make the massless action invariant under (3.13), these
transformations may not be a true invariance of the action, being at most a
symmetry of the equations of motion. If we compute the classical equation of
motion for $x^{\mu }$ that follows from the massive action (3.6) we will
find that 
\begin{equation}
\frac{d}{d\tau }(\frac{1}{\lambda }\dot{x}^{\mu })=0  \tag{3.14}
\end{equation}
This equation of motion is identical to the one that follows from the
massless action (3.8). We may say that at the classical level the massive
and the massless relativistic particles are governed by the same
relativistic dynamics. Now, while the classical equation of motion (3.14) is
invariant under transformation (3.13) when the ortogonality condition is
used, the massive action (3.6) is not. Transformation (3.13) is a symmetry
of the equations of motion only in the massive theory. In the massless
theory it is a symmetry of the equations of motion and of the action.

Now, invariance of the massless action under transformations (3.13) means
that infinitesimally we can define the scale transformation 
\begin{equation}
\delta x^{\mu }=\alpha \beta (\dot{x}^{2})x^{\mu }  \tag{3.15}
\end{equation}
where $\alpha $ is the same constant that appears in equations (2.4) and
(2.5). These transformations then lead to the existence of a new type of
dilatations. These new dilatations manifest themselves in the fact that the
vector field $D$ of equation (2.8) can be changed according to 
\begin{equation}
D=x^{\mu }\partial _{\mu }\rightarrow D^{\ast }=x^{\mu }\partial _{\mu
}+\beta (\dot{x}^{2})x^{\mu }\partial _{\mu }=D+\beta D  \tag{3.16}
\end{equation}
In fact, because all vector fields in equation (2.5) involve partial
derivatives with respect to $x^{\mu }$ , and $\beta $ is a function of $\dot{%
x}^{\mu }$ only, we can also introduce the generators 
\begin{equation}
P_{\mu }^{\ast }=P_{\mu }+\beta P_{\mu }  \tag{3.17}
\end{equation}
\begin{equation}
M_{\mu \nu }^{\ast }=M_{\mu \nu }+\beta M_{\mu \nu }  \tag{3.18}
\end{equation}
\begin{equation}
K_{\mu }^{\ast }=K_{\mu }+\beta K_{\mu }  \tag{3.19}
\end{equation}
and define the new space-time vector field 
\begin{equation}
V_{0}^{\ast }=a^{\mu }P_{\mu }^{\ast }-\frac{1}{2}\omega ^{\mu \nu }M_{\mu
\nu }^{\ast }+\alpha D^{\ast }+b^{\mu }K_{\mu }^{\ast }  \tag{3.20}
\end{equation}
The generators of this vector field obey the algebra 
\begin{equation}
\lbrack P_{\mu }^{\ast },P_{\nu }^{\ast }]=0  \tag{3.21a}
\end{equation}
\begin{equation}
\lbrack P_{\mu }^{\ast },M_{\nu \lambda }^{\ast }]=(\delta _{\mu \nu
}P_{\lambda }^{\ast }-\delta _{\mu \lambda }P_{\nu }^{\ast })+\beta (\delta
_{\mu \nu }P_{\lambda }^{\ast }-\delta _{\mu \lambda }P_{\nu }^{\ast }) 
\tag{3.21b}
\end{equation}
\begin{equation*}
\lbrack M_{\mu \nu }^{\ast },M_{\lambda \rho }^{\ast }]=(\delta _{\nu
\lambda }M_{\mu \rho }^{\ast }+\delta _{\mu \rho }M_{\nu \lambda }^{\ast
}-\delta _{\nu \rho }M_{\mu \lambda }^{\ast }-\delta _{\mu \lambda }M_{\nu
\rho }^{\ast })
\end{equation*}
\begin{equation}
+\beta (\delta _{\nu \lambda }M_{\mu \rho }^{\ast }+\delta _{\mu \rho
}M_{\nu \lambda }^{\ast }-\delta _{\nu \rho }M_{\mu \lambda }^{\ast }-\delta
_{\mu \lambda }M_{\nu \rho }^{\ast })  \tag{3.21c}
\end{equation}
\begin{equation}
\lbrack D^{\ast },D^{\ast }]=0  \tag{3.21d}
\end{equation}
\begin{equation}
\lbrack D^{\ast },P_{\mu }^{\ast }]=-P_{\mu }^{\ast }-\beta P_{\mu }^{\ast }
\tag{3.21e}
\end{equation}
\begin{equation}
\lbrack D^{\ast },M_{\mu \nu }^{\ast }]=0  \tag{3.21f}
\end{equation}
\begin{equation}
\lbrack D^{\ast },K_{\mu }^{\ast }]=K_{\mu }^{\ast }+\beta K_{\mu }^{\ast } 
\tag{3.21g}
\end{equation}
\begin{equation}
\lbrack P_{\mu }^{\ast },K_{\nu }^{\ast }]=2(\delta _{\mu \nu }D^{\ast
}-M_{\mu \nu }^{\ast })+2\beta (\delta _{\mu \nu }D^{\ast }-M_{\mu \nu
}^{\ast })  \tag{3.21h}
\end{equation}
\begin{equation}
\lbrack M_{\mu \nu }^{\ast },K_{\lambda }^{\ast }]=(\delta _{\lambda \nu
}K_{\mu }^{\ast }-\delta _{\lambda \mu }K_{\nu }^{\ast })+\beta (\delta
_{\lambda \nu }K_{\mu }^{\ast }-\delta _{\lambda \mu }K_{\nu }^{\ast }) 
\tag{3.21i}
\end{equation}
\begin{equation}
\lbrack K_{\mu }^{\ast },K_{\nu }^{\ast }]=0  \tag{3.21j}
\end{equation}
Notice that the vanishing brackets of the conformal algebra (2.10) are
preserved as vanishing in the above algebra, but the non-vanishing brackets
of the conformal algebra now have linear and quadratic contributions from
the arbitrary function $\beta (\dot{x}^{2})$. Algebra (3.21) is an off shell
extension of the conformal algebra (2.10).

Now consider the commutator structure induced by transformation (3.13a). We
assume the usual commutation relations between the canonical variables, $%
[x_{\mu },x_{\nu }]=[p_{\mu },p_{\nu }]=0$ , $[x_{\mu },p_{\nu }]=i\delta
_{\mu \nu }$. Taking $\beta (\dot{x}^{2})=\beta (\lambda ^{2}p^{2})$ in
transformation (3.13a) and transforming the $p_{\mu }$ in the same manner as
the $x_{\mu },$ we find that the new transformed canonical variables $(%
\tilde{x}_{\mu },\tilde{p}_{\mu })$ obey the commutators 
\begin{equation}
\lbrack \tilde{p}_{\mu },\tilde{p}_{\nu }]=0  \tag{3.22}
\end{equation}
\begin{equation}
\lbrack \tilde{x}_{\mu },\tilde{p}_{\nu }]=i\delta _{\mu \nu }(1+\beta
)^{2}+(1+\beta )[x_{\mu },\beta ]p_{\nu }  \tag{3.23}
\end{equation}
\begin{equation}
\lbrack \tilde{x}_{\mu },\tilde{x}_{\nu }]=(1+\beta )\{x_{\mu }[\beta
,x_{\nu }]-x_{\nu }[\beta ,x_{\mu }]\}  \tag{3.24}
\end{equation}
written in terms of the old canonical variables. These commutators obey the
non trivial Jacobi identities $(\tilde{x}_{\mu },\tilde{x}_{\nu },\tilde{x}%
_{\lambda })=0$ \ and \ $(\tilde{x}_{\mu },\tilde{x}_{\nu },\tilde{p}%
_{\lambda })=0$. They also reduce to the usual canonical commutators when $%
\beta (\lambda ^{2}p^{2})=0$.

The simplest example of this geometry is the case when $\beta (\lambda
^{2}p^{2})=\lambda ^{2}p^{2}$. The new positions then satisfy 
\begin{equation}
\lbrack \tilde{x}_{\mu },\tilde{x}_{\nu }]=-2i\lambda ^{2}M_{\mu \nu }^{\ast
}  \tag{3.25}
\end{equation}
where $M_{\mu \nu }^{\ast }$ is the extended off shell operator of Lorentz
rotations given by equation (3.18). The commutator (3.25) satisfies 
\begin{equation}
\int \mathbf{Tr}[\tilde{x}_{\mu },\tilde{x}_{\nu }]=0  \tag{3.26}
\end{equation}
as is the case for a general non-commutative algebra [8]. From equation
(3.25) we may say that there exists an inertial frame in which the
uncertainty introduced by two simultaneous position measurements is an off
shell rotation in space-time.

Finally we consider the case of the massless superparticle. It is described
by the action [9] 
\begin{equation}
S=\frac{1}{2}\int d\tau \lambda ^{-1}(\dot{x}^{\mu }-i\bar{\theta}\Gamma
^{\mu }\dot{\theta})^{2}  \tag{3.27}
\end{equation}
where $\theta _{\alpha }$ is a space-time spinor and $\Gamma _{\alpha \beta
}^{\mu }$ are Dirac matrices. The canonical momentum conjugate to $x^{\mu }$
is 
\begin{equation}
p_{\mu }=\frac{1}{\lambda }(\dot{x}_{\mu }-i\bar{\theta}\Gamma _{\mu }\dot{%
\theta})=\frac{1}{\lambda }Z_{\mu }  \tag{3.28}
\end{equation}
where we introduced the supersymmetric [9] variable $Z^{\mu }=\dot{x}^{\mu
}-i\bar{\theta}\Gamma ^{\mu }\dot{\theta}$. As in the bosonic case, the
primary constraint $p_{\lambda }=0$ leads to the mass shell condition $\phi =%
\frac{1}{2}p^{2}=0$.

If we extend the calculations in [7] to the case of the massless
superparticle we will find that the relation 
\begin{equation}
Z.\frac{dZ}{d\tau }=0  \tag{3.29}
\end{equation}
must hold. Equation (3.29) is the supersymmetric extension of the
special-relativistic condition of ortogonality between velocity and
acceleration. Again, condition (3.29) is an imposition of the supersymmetric
relativistic theory, and not an artificially introduced external ingredient.

Relaxing again the mass shell condition, we find that the massless
superparticle action (3.27) must be invariant under the transformation 
\begin{equation}
x^{\mu }\rightarrow \tilde{x}^{\mu }=\exp \{\beta (Z^{2})\}x^{\mu } 
\tag{3.30a}
\end{equation}
\begin{equation}
\theta _{\alpha }\rightarrow \tilde{\theta}_{\alpha }=\exp \{\frac{1}{2}%
\beta (Z^{2})\}\theta _{\alpha }  \tag{3.30b}
\end{equation}
\begin{equation}
\lambda \rightarrow \exp \{2\beta (Z^{2})\}\lambda  \tag{3.30c}
\end{equation}
where $\beta $ is now an arbitrary function of $Z^{2}$. In the canonical
formalism $\beta (Z^{2})=\beta (\lambda ^{2}p^{2}).$ Since the bosonic
momentum $p_{\mu }$ commutes with the fermionic canonical variables $\theta
_{\alpha }$ and $\pi _{\alpha }$ , this leaves invariant the
anti-commutation relations between the fermionic variables but change the
bosonic ones in the same way as (3.22-3.24). The commutator structure we
found for the bosonic massless particle is thus preserved in the
supersymmetric massless theory.

\end{document}